\documentclass[english,10pt,aps,prd,a4paper,preprintnumbers,floatfix,nofootinbib,showpacs,superscriptaddress]{revtex4-1}
\usepackage[english]{babel}
\usepackage{amsmath}
\usepackage{graphicx}
\usepackage{color}
\usepackage{mathrsfs}   
\usepackage{amssymb}
\usepackage{hyperref}
\usepackage{dcolumn}% Align table columns on decimal point
\usepackage{bm}% bold math
\usepackage{graphicx}
\usepackage[sort&compress]{natbib}
\newcommand {\Tr} {{\rm Tr}}
\newcommand {\be} {\begin{equation}}
\newcommand {\ee} {\end{equation}}

\definecolor{greenLinks}{rgb}{0, 0.6, 0} 
\definecolor{blueLinks}{rgb}{0, 0, 0.6}
\definecolor{redLinks}{rgb}{0.6, 0, 0}
\definecolor{tempText}{rgb}{0.55, 0.10,0.67}
\definecolor{eprintLinks}{rgb}{0.4, 0.4, 0.4}
\definecolor{journalLinks}{rgb}{0.6, 0, 0}

\def\vev#1{\left\langle #1\right\rangle}

\usepackage{float}

\def\vev#1{\left\langle #1\right\rangle}

\def\21{$\mathrm{SU(2)_L \otimes U(1)_Y}$ }
\def\31{$\mathrm{SU(3)_c \otimes U(1)_Q}$ }

\def\3211{$\mathrm{SU(3) \otimes SU(2)_L \otimes U(1)_R \otimes U(1)_{B-L}}$ }
\def\321{$\mathrm{SU(3) \otimes SU(2) \otimes U(1)}$ }
\def\422{$\mathrm{SU(4) \otimes SU(2) \otimes SU(2)_R}$ }
\newcommand {\ignore}[1]{}

\newcommand{\sm}{{Standard Model }}

\def\vev#1{\left\langle #1\right\rangle}

\def\U1{$\mathrm{ U(1)_{B_3 - 3 L_\mu} }$}
\def\vev#1{\left\langle #1\right\rangle}

\newcommand{\AddrUNAM}{ {\it Instituto de F\'{\i}sica, Universidad Nacional Aut\'onoma de M\'exico (UNAM), A.P. 20-364, Ciudad de M\'exico 01000, MEXICO.}}
\newcommand{\AddrAHEP}{%
  AHEP Group, Institut de F\'{i}sica Corpuscular --
  C.S.I.C./Universitat de Val\`{e}ncia, Parc Cient\'ific de Paterna.\\
 C/ Catedr\'atico Jos\'e Beltr\'an, 2 E-46980 Paterna (Valencia) - SPAIN}

\begin{document}
%\title{Type-II seewsaw for Dirac neutrinos with flavour}
\title{Flavour--symmetric type-II Dirac neutrino seesaw mechanism}
\author{Cesar Bonilla}\email{cesar.bonilla@ific.uv.es} \affiliation{\AddrAHEP}
\author{J. M. Lamprea}\email{jmlamprea@estudiantes.fisica.unam.mx}\affiliation{\AddrUNAM}
\author{Eduardo Peinado} \email{epeinado@fisica.unam.mx}\affiliation{\AddrUNAM}
\author{Jose W. F. Valle}\email{valle@ific.uv.es}\affiliation{\AddrAHEP}

\begin{abstract}
  \vspace{1cm} We propose a Standard Model extension with underlying
  $A_4$ flavour symmetry where small Dirac neutrino masses arise from
  a Type--II seesaw mechanism. The model predicts the ``golden''
  flavour-dependent bottom-tau mass relation, requires an inverted
  neutrino mass ordering and non-maximal atmospheric mixing
  angle. Using the latest neutrino oscillation global
  fit~\cite{deSalas:2017kay} we derive restrictions on the oscillation
  parameters, such as a correlation between $\delta_{CP}$ and
  $m_{\nu_\text{lightest}}$.
\end{abstract}

\maketitle

\section{Introduction}

Full quark-lepton correspondence within the \sm would suggest
neutrinos to be Dirac fermions, with the lepton mixing matrix
completely analogous to the CKM matrix.
However, if the ultimate description of particle physics is a
four-dimensional quantum field theory, this situation is unlikely
since, on general grounds, one expects neutrinos to be
Majorana type~\cite{Schechter:1980gr}.
In addition, the associated mechanisms to account for small neutrino
mass as a consequence of their charge neutrality, such as the
celebrated seesaw mechanism in its various
realisations~\cite{Schechter:1980gr,Minkowski:1977sc,GellMann:1980vs,mohapatra:1980ia,Lazarides:1980nt},
all lead to Majorana neutrinos.

From the experimental side, however, despite great efforts over
several decades, neutrinoless double beta decay has not yet been
detected~\cite{KamLAND-Zen:2016pfg}.  According to the black-box
theorem~\cite{Schechter:1981bd,Duerr:2011zd}, such detection would
provide the only robust way to establish the Majorana nature of
neutrinos.
Hence, for the time being, one must keep an open mind and re-examine
the foundations of our ideological prejudices against having neutrinos
as Dirac fermions.

Till recently, one argument against Dirac neutrinos has been the
absence of convincing realizations of the seesaw mechanism for this
case.
However this argument is fragile, and Dirac seesaw mechanisms have
been shown to exist, both within the
type-I~\cite{Ma:2015raa,Chulia:2016ngi,CentellesChulia:2017koy}, as
well as type-II~\cite{Bonilla:2016zef,Valle:2016kyz,Reig:2016ewy}
realizations, for a brief classification see~\cite{
  Ma:2016mwh}~\footnote{For the possibility of having radiative Dirac
  neutrino mass models
  see~Refs.~\cite{Bonilla:2016diq,Borah:2017leo,Wang:2017mcy}.}.
Moreover, the existence of right-handed states, required for Dirac
masses, may be necessary in order to have a consistent high energy
completion, or for realizing a higher symmetry, such as the gauged B-L
symmetry present in the conventional SO(10) seesaw scenarios.

Dirac neutrinos also arise in schemes with extra space-time dimensions
such as string constructions, where right-handed states are needed to
ensure anomaly cancellation through a generalized Green-Schwarz
mechanism~\cite{Addazi:2016xuh}.
Alternatively, Dirac neutrinos arise from five-dimensional anti de
Sitter space warped solutions to the hierarchy
problem~\cite{Chen:2015jta}.
The truth of the matter is that the nature of neutrinos remains as
mysterious as the mechanism providing their small masses, and hence we
can not simply rule it out.

In this letter, we consider the possibility of Dirac neutrinos
resulting from a family symmetry construction in which the small
neutrino mass arises {\it \`{a} la seesaw}, thus complementing the
idea proposed in~\cite{Aranda:2013gga}.
In addition to shedding light upon the pattern of neutrino
oscillations, we also require the flavour structure to provide the
generalised bottom-tau mass relation,
\begin{equation}
\frac{m_\tau}{\sqrt{m_e m_\mu}} = \frac{m_b}{\sqrt{m_d m_s}},
\label{Eq:MasRel}
\end{equation}
proposed in~\cite{Morisi:2009sc,Morisi:2011pt}. This successful
``golden'' mass relation is derived without the need for invoking
grand unification in a number of
models~\cite{CentellesChulia:2017koy,King:2013hj,Morisi:2013eca,Bonilla:2014xla,Carballo-Perez:2016ooy},
all of which lead to particular restrictions on the neutrino
oscillation sector.
The paper is organized as follows: In the next section we introduce
the model, in section~\ref{sec:flav-pred-numer} we present the
predictions and discuss the results, and finally we conclude
in~\ref{Sec:Con}.

\section{The model}
\label{sec:model}

\begin{figure}[h!]
\begin{center}
\includegraphics[scale=0.3]{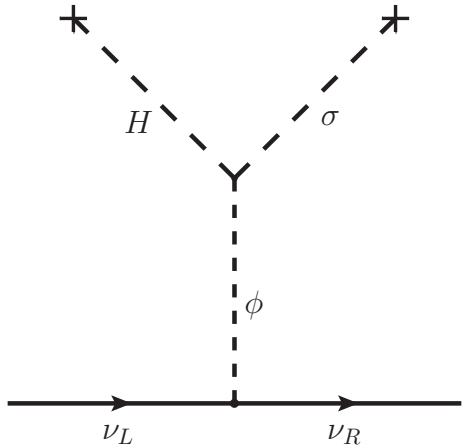}
\caption{{\it Neutrino mass generation in the Type-II seesaw for Dirac
    neutrinos~\cite{Bonilla:2016zef,Valle:2016kyz,Reig:2016ewy}}\label{fig:FeyDiag}}
\end{center}
\end{figure}

This model is constructed so that the smallness of the neutrino mass
has a dynamical origin, given by the Type-II seesaw mechanism for
Dirac neutrinos, illustrated in Figure~\ref{fig:FeyDiag}. 
In order to explain neutrino oscillation data in this scenario we
introduce the flavour symmetry
$A_4 \otimes \mathbb{Z}_3 \otimes \mathbb{Z}_2$.
After symmetry breaking there is an Abelian discrete symmetry
$\mathbb{Z}_3$ which forbids the appearance of Majorana operators at
the loop level in the broken phase, thus protecting the Diracness of
neutrinos.

\subsection{Lepton sector}
\label{sec:lepton-sector}

The particle assignments for the lepton sector and scalars are given
in Table~\ref{Tab:Mod}.
%% %
%The vacuum expectation value (vev) alignments of these scalar triplets
%given by $ \vev{H^d} = (v_{h^d_1}, v_{h^d_2}, v_{h^d_3})$ and
%$\vev{\phi} = (v_{\phi_1}, v_{\phi_2}, v_{\phi_3})$. The complex
%scalar $\sigma$ responsible for inducing the small vev of $\phi$ could
%transform either as triplet or singlet under $A_4$. 
%
%On the other hand all leptons, both left- and right-handed, including
%the right-handed neutrinos $\nu_R =(\nu_{R_1},\nu_{R_2},\nu_{R_3})$,
%transform as $A_4$ triplets.
%
The $SU(2)$ scalar doublets $H^d = (H^d_1, H^d_2, H^d_3)$ and
$\phi = (\phi_1, \phi_2, \phi_3)$ transform as triplets under $A_4$,
where each component can be written as follows
\begin{equation}
H^{d}_i = 
\begin{pmatrix}
h^{d \,+}_i \\
h^{d\,0}_i
\end{pmatrix}, 
\qquad
\phi_i = 
\begin{pmatrix}
\phi^0_i \\
\phi^-_i
\end{pmatrix}.
\label{Eq:Higgses}
\end{equation}

\begin{table}[h!]
\begin{center}
\begin{tabular}{| c | c | c | c | c | c | c | }
\hline
 & $\bar{L}$  & $\ell_R$ & $\nu_R$ & $H^d$ & $\phi$ & $\sigma$\\
\hline
\hline
$SU(2)_L \otimes U(1)_Y$ & $(2,-1/2)$ & $(1,-1)$ & $(1,0)$ & $(2,1/2)$ & $(2,-1/2)$ & $(1,0)$\\
%$Y$ & -1/2 & -1 & 0 &1/2 &-1/2 & 0\\
\hline
$A_4$ & {\bf 3} & {\bf 3} & {\bf 3} & {\bf 3} & {\bf 3} & {\bf 3} or ${\bf 1_i}$\\
\hline
$\mathbb{Z}_3$ & $\omega^2$ & $\omega$ & $\omega$ &  1 &  1 &  1 \\
\hline
$\mathbb{Z}_2$ & $+$ & $+$ & $-$ & $+$ & $-$ & $-$  \\
\hline
\end{tabular}
\caption{\it Charge assignments for the particles involved in the neutrino mass 
generation mechanism,  where $\omega^3=1$.} 
\label{Tab:Mod}
\end{center}
\end{table}

The vacuum expectation value (vev) alignments of these scalar triplets
given by $ \vev{H^d} = (v_{h^d_1}, v_{h^d_2}, v_{h^d_3})$ and
$\vev{\phi} = (v_{\phi_1}, v_{\phi_2}, v_{\phi_3})$. The complex
scalar $\sigma$ responsible for inducing the small induced vev of
$\phi$ could transform either as triplet or singlet under $A_4$. On
the other hand all leptons, both left- and right-handed, including the
right-handed neutrinos $\nu_R =(\nu_{R_1},\nu_{R_2},\nu_{R_3})$,
transform as $A_4$ triplets.
Given the charges under the cyclic groups
$\mathbb{Z}_3\otimes \mathbb{Z}_2$, one can easily see that the
$\mathbb{Z}_3$ remains unbroken after symmetry breaking, because all
scalars are blind under this symmetry. 
Such residual $\mathbb{Z}_3$ symmetry forbids the Majorana mass terms
$M_R\, \nu_R\,\nu_R$ as well as the dimension-5 operators:
$L H^d L H^d$, $L \tilde{\phi} L \tilde{\phi}$ and $LH^d L \tilde{\phi}$.
This symmetry also forbids higher order operators derived from the
product of the previous dimension-5 operators and
$(H^{d\,\dagger} H^d)^n$, $(\phi^\dagger \phi)^n$ and
$(H^{d \dagger} \tilde{\phi})^n$ as well as $ \nu_R\,\nu_R\,\sigma^n$.
Finally, the $\mathbb{Z}_2$ charges are assigned as $(\bm{-})$ to
$\nu_R$, $\sigma$ and $\phi$ and $+$ to the other particles.
As a result they act in a complementary way to the $\mathbb{Z}_3$,
forbiding the unwanted renormalisable Yukawa couplings:
$\bar{L}\, \tilde{\phi}\, \ell_R$ and $\bar{L}\,\tilde{H}^d\, \nu_R$,
where $\tilde{\phi} = i \sigma_2 \phi^* $ and
$\tilde{H^d} = i \sigma_2 H^{d\,*}$.

In accordance with the previous discussion, the relevant part of
Yukawa Lagrangian for the leptons is given as:
\begin{equation}
\mathcal{L}_Y \supset Y_\ell^i \, \left[ \bar{L}, H^d\right]_{3_i} \ell_R + Y_\nu^i \, \left[ \bar{L}, \phi \right]_{3_i}  \nu_R + \text{h.c.},
\label{Eq:YukLag} 
\end{equation}
where the symbol $[ a , b ]_{3_i}$ stands for the two ways of
contracting two triplets of $A_4$, $a$ and $b$, into a triplet, as
shown in the Appendix.
Before proceeding we summarize our model structure by saying that,
compared with the minimal \sm case, here one has an extra scalar
iso-doublet $\phi$, the iso-singlet $\sigma$ and the right-handed
neutrino states $\nu_R$, all triplets under $A_4$.  After symmetry
breaking, neutrinos get a small type-II seesaw mass, as a result of
the small vacuum expectation value (vev) $\vev{\phi}$ which is induced
by the vev of $\sigma$ as proposed
in~\cite{Bonilla:2016zef,Valle:2016kyz,Reig:2016ewy}.

\section{Flavour predictions and numerical results}
\label{sec:flav-pred-numer}

Two aspects of the flavour problem concern the explanation of mass
hierarchies of quark and leptons, as well as to account for the
structure of mixing in each of these sectors, so disparate from each
other.
We will see how our model leads to a successful ``golden'' mass
formula relating quark and lepton masses, despite the absence of grand
unification. 
This is a flavour generalization of bottom-tau unification previously
proposed
in~\cite{CentellesChulia:2017koy,Morisi:2011pt,King:2013hj,Morisi:2013eca,Bonilla:2014xla,Carballo-Perez:2016ooy}.
In addition we will derive the corresponding specific predictions for
the lepton mixing matrix describing neutrino oscillations. This arises
from the study of the charged and neutral lepton sector.
While no predictions are made for the CKM quark mixing matrix, it can
be adequately fit in a simple way, see reference~\cite{King:2013hj}.
We now spell out the detailed flavour predictions of our model.

\subsection{Charged fermions and the generalised 
bottom-tau mass relation}
\label{sec:charg-ferm-gener}

The complete particle assignment of our model is inspired in the one
in~\cite{King:2013hj}, and is shown in Table \ref{Tab:ModCom}, including
both gauge as well as flavour transformation properties.
\begin{table}[h!]
\begin{center}
\begin{tabular}{| c || c | c | c | c | c | c || c | c | c | c |}
\hline
 & $\bar{Q}$ & $\bar{L}$  & $u_R$ & $d_R$ &$\ell_R$ & $\nu_R$ & $H^u_i$ & $H^d$ & $\phi$ & $\sigma$ \\
\hline
\hline
$SU(2)_L \otimes U(1)_Y$ & $(2,1/6)$ & $(2,-1/2)$ & $(1,2/3)$ & $(1,-1/3)$ & $(1,-1)$ & $(1,0)$ & $(2,-1/2)$ & $(2,1/2)$ & $(2,-1/2)$ & $(1,0)$ \\
%\hline
%$Y$ & $1/6$ & $-1/2$ & $2/3$ & $-1/3$ & -1 & 0 & $1/2$ &$1/2$ & $-1/2$ & 0 \\
\hline
$A_4$ & {\bf 3} & {\bf 3} & ${\bf 1_i}$ & {\bf 3} & {\bf 3} & {\bf 3} & ${\bf 3}$ & {\bf 3} & {\bf 3}  & {\bf 3}/${\bf 1_i}$\\
\hline
$Z_3$ & $1$ & $\omega^2$ & $1$ & $1$ & $\omega$ & $\omega$ & $1$ & $1$ & $1$ & $1$ \\
\hline
$Z_2$ & $+$ & $+$ & $+$ & $+$ & $+$ & $-$ & $+$ & $+$ & $-$ & $-$ 
\\
\hline
$Z_2^d$ & $+$ & $+$ & $+$ & $-$ & $+$ & $+$ & $+$ & $-$ & $+$ & $+$ \\
\hline
\end{tabular}
\caption{\it Particle content and quantum numbers for the model.}
\label{Tab:ModCom}
\end{center}
\end{table}

From Eq. (\ref{Eq:YukLag}) one sees that the charged lepton mass
matrix, following~\cite{Morisi:2011pt,King:2013hj,Morisi:2013eca}, can
be parametrised as:
\begin{equation}
m_\ell =
\begin{pmatrix}
0 & a_\ell \alpha_\ell e^{i \theta_\ell} & b_\ell \\
b_\ell \alpha_\ell & 0 &  e^{i \theta_\ell} a_\ell \rho_\ell \\
a_\ell  e^{i \theta_\ell}  & b_\ell \rho_\ell & 0 
\end{pmatrix},
\end{equation}
where $a_\ell= v_{h^d_2} (Y_\ell^1 + Y_\ell^3) $ and
$b_\ell = v_{h^d_2} (Y_\ell^2 + Y_\ell^4) $ are real Yukawa couplings,
$\theta_\ell$ is a unremovable complex phase\footnote{ We
    assume that the non-conservation of CP symmetry comes entirely
    from the neutrino sector.  Thus in our analysis we fixed
    $\theta_\ell$=0.} and the $H^d$ vev alignment is parameterised as
$ \vev{H^d} = (v_{h^d_1},v_{h^d_2},v_{h^d_3}) = v_{h^d_2} (\rho_\ell,
1, \alpha_\ell)$,
with $ \alpha_\ell = v_{h^d_3}/v_{h^d_2}$ and
$\rho_\ell= v_{h^d_1}/v_{h^d_2}$.

The bi--unitary invariants of the squared mass matrix
$M^2_\ell = m_\ell m_\ell^\dagger$ are determined as:
\begin{align}
\Tr M^2_\ell &= m_1^2 +m_2^2+ m_3^2,\label{Eq:Inv1}\\
\det M^2_\ell &= m_1^2 m_2^2 m_3^2,\label{Eq:Inv2}\\
(\Tr M_\ell^2)^2 -\Tr(M_\ell^2)^2 &= 2 m_1^2 m_2^2 + 2 m_2^2 m_3^2 + 2 m_1^2 m_3^2 \label{Eq:Inv3}.
\end{align}
We work under the assumptions $\rho_\ell \gg \alpha_\ell$, $\rho_\ell \gg 1$, $b_\ell>a_\ell$ and
$\rho_\ell \gg \frac{b_\ell}{a_\ell}$ which, at leading order, ensure adequate family
mass hierarchy as well as mixing patterns. One can show from
Eqs. (\ref{Eq:Inv1}--\ref{Eq:Inv3}) that:
\begin{align}
(b_\ell \rho_\ell)^2 &\approx m_3^2,\label{Eq:InvApp1}\\
(b_\ell^3 \rho_\ell \alpha_\ell)^2  &\approx m^2_1 m^2_2 m^2_3,\label{Eq:InvApp2}\\
(a_\ell b_\ell \rho_\ell^2)^2 &\approx m^2_2 m_3^2\label{Eq:InvApp3}.
\end{align}
Solving the system in Eqs.~(\ref{Eq:InvApp1}-\ref{Eq:InvApp3}), we can
find the approximate expressions:
\begin{equation}
\label{Eq:SolPar}
a_\ell \approx \frac{m_2}{m_3} \sqrt{\frac{m_1 m_2}{\alpha_\ell}},  \qquad b_\ell \approx 
\sqrt{\frac{m_1 m_2}{\alpha_\ell}} \qquad \text{and} \qquad  \frac{\rho_\ell}{\sqrt{\alpha_\ell}} \approx  \frac{m_3}{\sqrt{m_1 m_2}}.
\end{equation}

Notice that also the down--type quarks couple to the $H^d$ and hence
have the same flavour structure. This implies that the parameters
$\rho_\ell$ and $\alpha_\ell$ in Eq.~(\ref{Eq:SolPar}) are common to
the charged leptons and the down--type quarks, i.e.
$\rho_\ell=\rho_d$ and $\alpha_\ell=\alpha_d$.
From this we derive the generalised bottom--tau mass relation in Eq. (\ref{Eq:MasRel}):
\[
\frac{m_\tau}{\sqrt{m_e m_\mu}} = \frac{m_b}{\sqrt{m_d m_s}},
\]%
in a straightforward way. This generalised down quark--charged lepton
mass relation, Eq. (\ref{Eq:MasRel}), follows from our flavour group
assignments.
Although it has been obtained also in other realizations of $A_4$
family symmetry~\cite{Morisi:2009sc,King:2013hj,Morisi:2013eca}, these
are not equivalent. 

\subsection{Fermion masses and mixing}
\label{sec:dirac-neutr-mass}

In this section we focus on the lepton mixing matrix, because it is in
this sector that our model makes non-trivial predictions. However the
CKM matrix describing quark mixing will be adequately described and
this provides an input for the lepton mixing matrix.
In analogy with the previous subsection, Eq. (\ref{Eq:YukLag}) gives
the neutrino mass matrix, which can also be parametrised as:
\begin{equation}
m_\nu =
\begin{pmatrix}
0 & a_\nu \alpha_\nu & b_\nu e^{i \theta_\nu} \\
b_\nu e^{i \theta_\nu} \alpha_\nu & 0 & a_\nu \rho_\nu \\
a_\nu  & b_\nu e^{i \theta_\nu} \rho_\nu & 0 
\end{pmatrix},
\label{Eq:MassNeu}
\end{equation}
where $a_\nu = v_{\phi_2} (Y_\nu^1 + Y_\nu^3)$ and
$b_\nu = v_{\phi_2}(Y_\nu^2+ Y_\nu^4)$ are real Yukawa couplings,
$\theta_\nu$ is the complex phase that cannot be rotated away under
$SU(2)$ transformations, and characterizes the strentgh of CP
violation in the lepton sector.
The vev--alignment of $\phi$ can be written as
$ \vev{\phi} = ( v_{\phi_1},v_{\phi_2},v_{\phi_3}) = v_{\phi_2} (
\rho_\nu ,1, \alpha_\nu) $,
with $ \alpha_\nu = v_{\phi 3}/v_{\phi 2}$ and
$\rho_\nu= v_{\phi 1}/v_{\phi 2}$.\\[-.2cm]

From the invariants Eqs.(\ref{Eq:Inv1}--\ref{Eq:Inv3}) of the squared
neutrino mass matrix $M_\nu^2 = m_\nu m_\nu^\dagger$:
\begin{align}
\Tr(M_\nu^2) &= (a_\nu^2 + b_\nu^2)(1+\alpha_\nu^2+\rho_\nu^2),\label{Eq:InvNu1}\\
\det(M_\nu^2) &= (a_\nu^6 + b_\nu^6 + 2 a_\nu^3 b_\nu^3 \cos (3\theta_\nu))\alpha_\nu^2 \rho_\nu^2, \label{Eq:InvNu2}\\
\frac{1}{2} \left[ (\Tr{M_\nu^2})^2 -\Tr({M_\nu^4})\right]  &=  a_\nu^2 b_\nu^2 (1+\alpha_\nu^4 + \rho_\nu^4) + (a_\nu^4+b_\nu^4) (\rho_\nu^2 + \alpha_\nu^2(1+\rho_\nu^2)) \label{Eq:InvNu3},
\end{align}
we performed a numerical scan over the parameter regions for the
solutions of Eqs. (\ref{Eq:InvNu1}--\ref{Eq:InvNu3}) that reproduce
the measured elements of the leptonic mixing matrix
$V = U_l^\dagger U_\nu$. To do this we use as inputs the $3\sigma$
values for the three neutrino mixing angles and the two mass squared
differences from the global fit~\cite{deSalas:2017kay}. The $U_\nu$
matrix comes from the bi--unitary transformation for the neutrinos
$ U^\dagger_\nu m_\nu V_\nu = D_\nu$, where
$D = \text{diag} \left(m_{\nu_1}, m_{\nu_2}, m_{\nu_3}\right)$, while
the $U_l$ comes from the equivalent transformation for the charged
leptons, e.g. $ U^\dagger_\ell m_\ell V_\ell = D_\ell$. \\[-.2cm]

We now turn to the CKM matrix describing quark mixing.
 Although we have no family symmetry prediction for the CKM matrix, we
 notice that it can be accomodated in the same way as described
 in~\cite{King:2013hj}.
 This fixes the value for the $\alpha_d$ parameter which enters also
 in the leptonic sector.
 In order to adequately fit the CKM matrix we need
 $\alpha_d=\alpha_l \approx 1.58$~\cite{King:2013hj}.
 Taking into account the numerical values for the charged lepton
 masses $m_e = 0.511006 ~\text{MeV}$, $m_\mu = 105.656 ~\text{MeV} $
 and $m_\tau = 1776.96 ~\text{MeV} $ and assuming that the complex
 phase $\theta_\ell = 0$, as in
 Refs.~\cite{Morisi:2011pt,King:2013hj,Morisi:2013eca}, we find that
 the resulting contribution from the charged lepton sector to the
 neutrino mixing matrix is fixed and close to be diagonal, see for
 instance~\cite{Morisi:2011pt}.

\subsection{Neutrino oscillation predictions}
\label{sec:neutr-oscill-pred}

In order to determine the neutrino oscillation predictions of the
model, we have performed a numerical scan in which parameters are
varied randomly in the ranges 
\begin{eqnarray}
\alpha_\nu \in \left[-10,10\right], \ \ 
\rho_\nu \in \left[-10,10\right]\ \ \text{and} \ \ \theta_\nu \in 
\left[0,2\pi\right].   
\end{eqnarray}

\begin{figure}[h!]
\begin{center}
\includegraphics[scale=0.5]{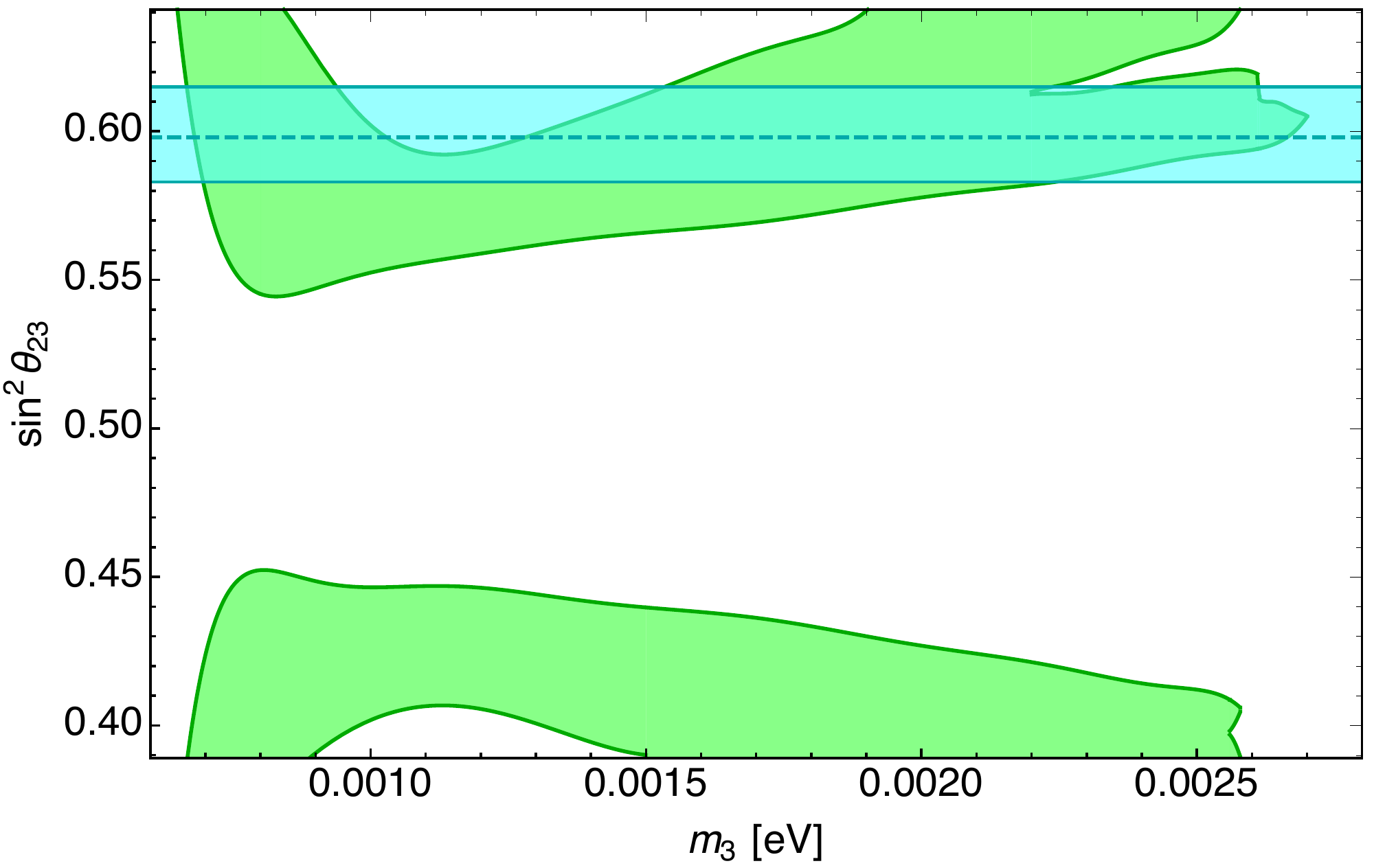}
\caption{{\it The regions in the atmospheric mixing angle
    $\theta_{23}$ and the lightest neutrino mass $m_3$ plane allowed
    by current oscillation data are the shaded (green) areas, see
    text. The horizontal dashed line represents the best-fit value
    for $\sin^2 \theta_{23}$, whereas the horizontal shaded region
    corresponds to the $1\sigma$ allowed region from
    Ref.~\cite{deSalas:2017kay}.} \label{fig:Sin32vsM3}}
\end{center}
\end{figure}

\begin{figure}[h!]
\begin{center}
\includegraphics[scale=0.41]{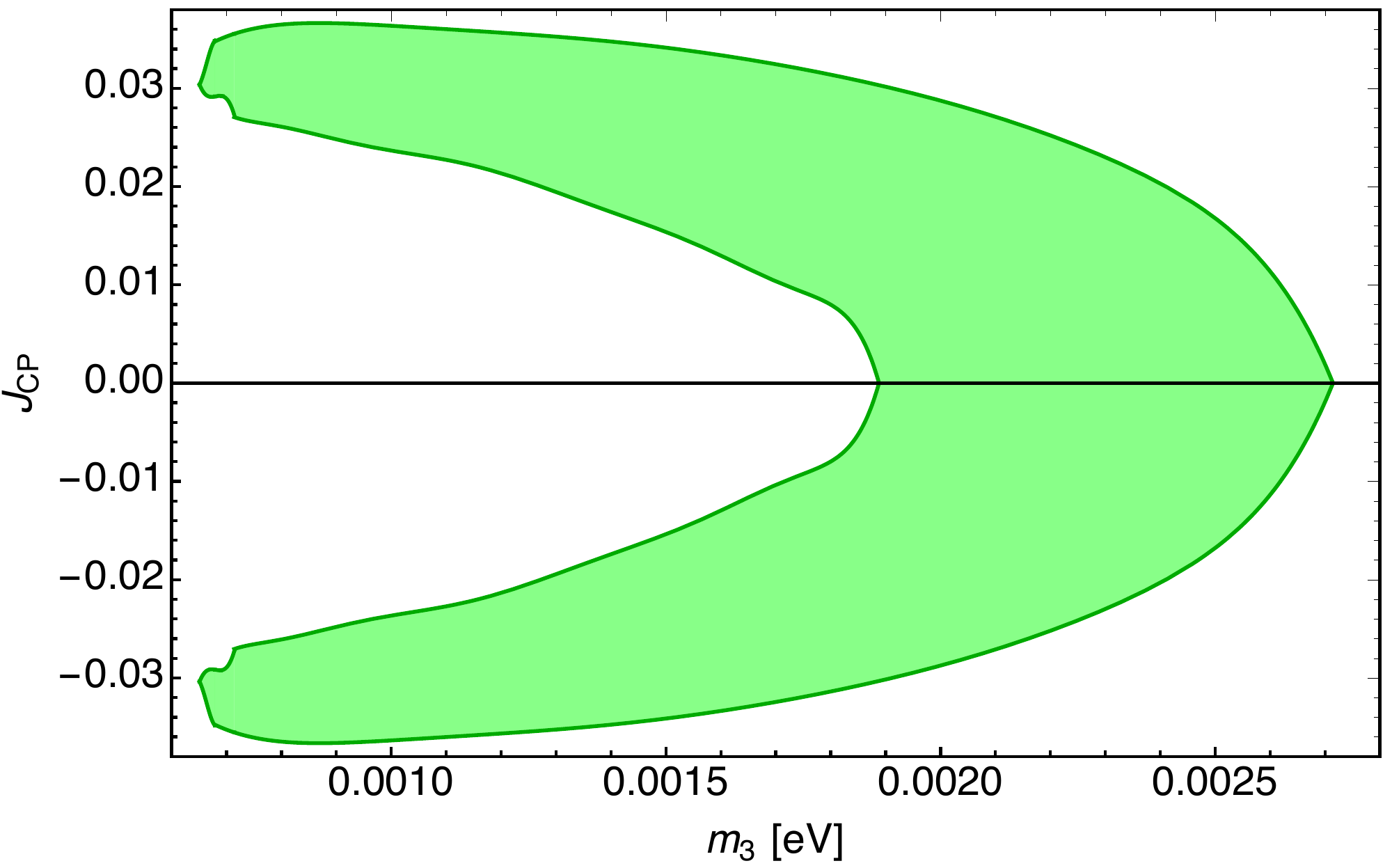}~~~\includegraphics[scale=0.4]{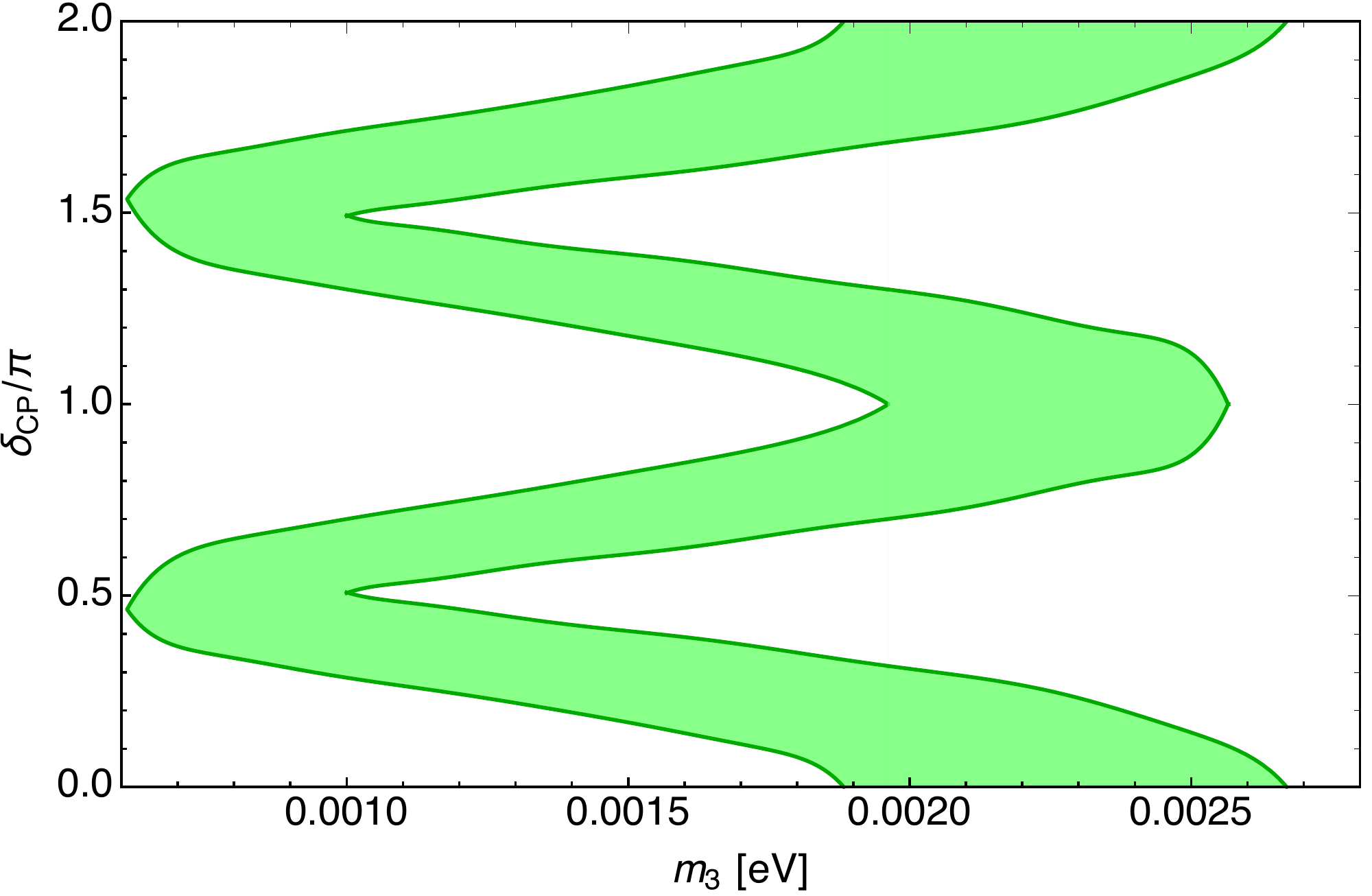}
\caption{{\it Correlation between the CP violation and the lightest neutrino mass. Left: correlation between the Jarlskog invatiant and the lightest neutrino mass  $m_3$ allowed by the current oscillation data from  Ref.~\cite{deSalas:2017kay}. Right: We plot also the allowed region for the correlation between the 
     the Dirac CP phase $\delta_{CP}$ and the lightest neutrino mass  $m_3$.}\label{fig:SM2vsJ}}
\end{center}
\end{figure}

\begin{figure}[h!]
\begin{center}
\includegraphics[scale=0.45]{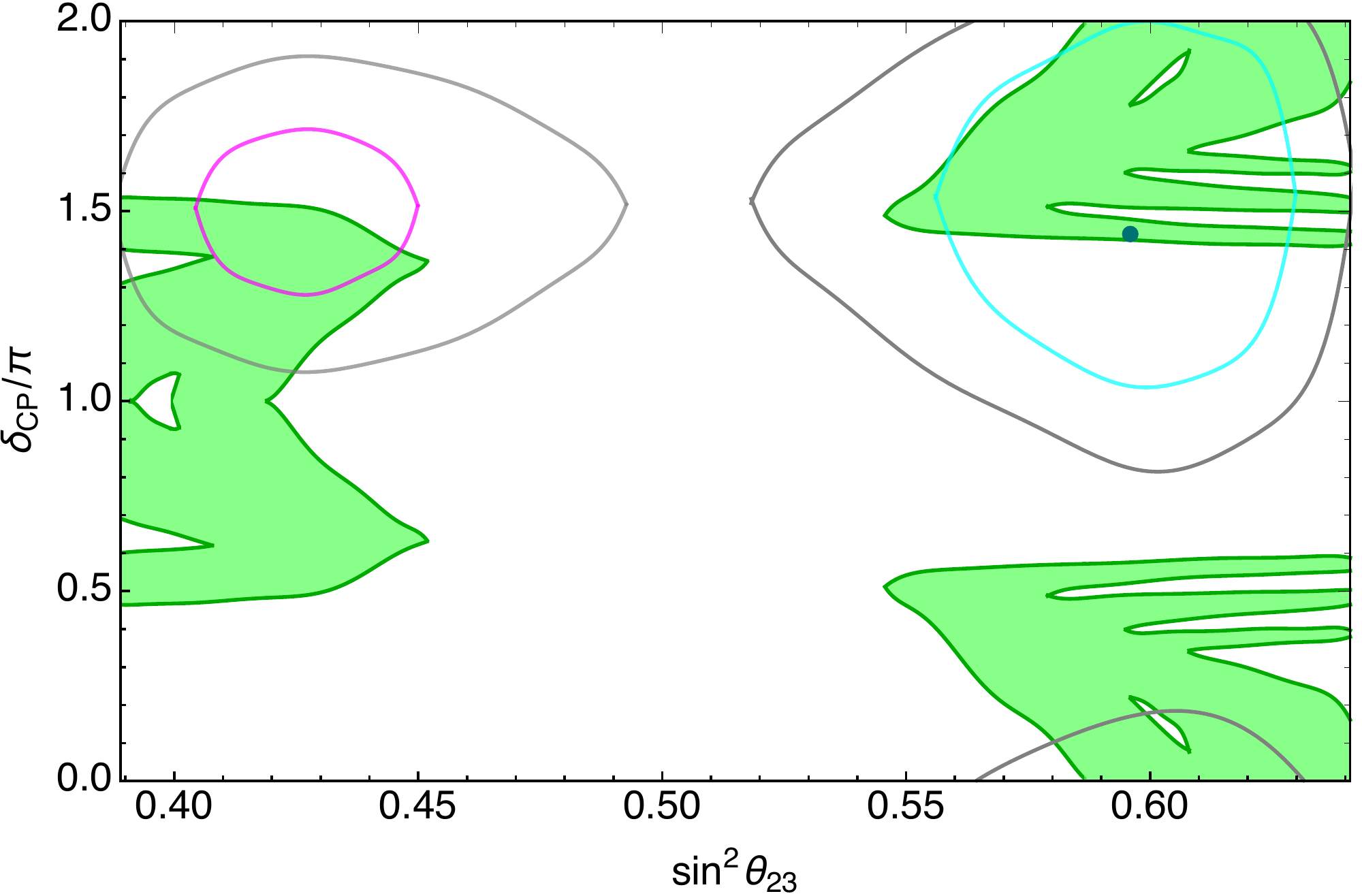}
\caption{{\it The allowed regions of the atmospheric mixing angle and
    $\delta_{CP}$ are indicated in shaded (green). They result from a
    numerical scan keeping only those choices that lie within
    3$\sigma$ of their preferred best fit values
    Ref.~\cite{deSalas:2017kay}} The unshaded regions are 90 and
  99\%CL regions obtained directly in the unconstrained
  three--neutrino oscillation global fit~\cite{deSalas:2017kay}.}
\label{Fig:Sin23VsDCP}
\end{center}
\end{figure}

Only those choices for which the undisplayed and well-measured
oscillation parameters are within 3$\sigma$ of the values obtained in
the latest neutrino oscillation global fit of
Ref.~\cite{deSalas:2017kay} are kept.
This way we have obtained the model--allowed regions in terms of the
``interesting'' and poorly determined oscillation parameters
$\theta_{23}$, $\delta_{CP}$ as well as the lightest neutrino mass
eigenvalue. 
  These are displayed in shaded (green) regions in the Figures
  \ref{fig:Sin32vsM3}, ~\ref{fig:SM2vsJ} and ~\ref{Fig:Sin23VsDCP}.
  In contrast, the unshaded regions are the 90 and 99\%CL regions
  obtained directly from the unconstrained three--neutrino oscillation
  global fit~\cite{deSalas:2017kay}, see Figures 4 and 5 in
  reference~\cite{deSalas:2017kay}.

We find that the model is only compatible with the inverted ordering
for the neutrino mass eigenvalues.  
The consistent parameter regions for the atmospheric mixing angle
$\sin \theta_{23}$ vs. the lightest neutrino mass $m_3$ are given in
Fig.~\ref{fig:Sin32vsM3}, while the CP violating Dirac phase
$\delta_{CP}$ vs. $m_3$ are displayed in Fig.~\ref{fig:SM2vsJ}.

From the plot given in Fig.~\ref{fig:SM2vsJ} one sees that the allowed
region for the lightest neutrino mass $m_3$ is within the range
$[6.4\times 10^{-4}~\mbox{eV},2.7 \times10^{-3}~\mbox{eV}]$. We can see
that only masses above $\sim 0.002$ eV allow $\delta_{CP} = 0$,
signifying no CP violation, while for lower masses such value is
always non--zero.
The shaded areas in Fig.~\ref{Fig:Sin23VsDCP} are obtained from a
numerical scan that filters those parameter choices for which the
well-measured undisplayed oscillation parameters lie within 3$\sigma$
of the best fit values obtained in the latest neutrino oscillation
global fit in Ref.~\cite{deSalas:2017kay}.
These should be compared with the unshaded 90 and 99\%CL regions
obtained directly in the unconstrained three--neutrino oscillation
global fit~\cite{deSalas:2017kay}.

\section{Conclusions}
\label{Sec:Con}

We have proposed a $A_4 \otimes \mathbb{Z}_3 \otimes \mathbb{Z}_2$
flavour extension of the Standard Model where the small neutrino
masses are generated from a type II Dirac see-saw mechanism. The model
addresses both aspects of the flavour problem: the explanation of mass
hierarchies of quark and leptons, as well as restricting the structure
of the lepton mixing matrix.
Concerning the first point, our model leads to a successful ``golden''
mass relation between quark en lepton masses, proposed previously.
In addition, the model provides flavour predictions for the lepton
mixing matrix relevant in order to account for neutrino oscillations.
First of all, inverted neutrino mass ordering and non-maximal
atmospheric mixing angle are predicted. 
While this is at odds with the results of the latest neutrino
oscillation global fit in~\cite{deSalas:2017kay}, we stress that the
neither the preference for normal ordering nor the indication for a
given octant are currently statistically significant, since the
general three-neutrino fit gives four possible closely separated local
minima.
In any case, taken at face value, the model at hand would suggest a
slight preference for the higher octant, since it predicts inverted
neutrino mass ordering.
We have also found a positive hint for CP violation, $\delta_{CP}\neq0$,
if $m_{\nu_\text{lightest}} \lesssim 0.002$~eV, while bigger masses
are consistent with CP conserving solutions.
Concerning the CKM quark mixing matrix, we also saw that, although no
definite predictions are made, the required CKM matrix elements can be
adequately described, and they also fix the contribution to the
neutrino mixing matrix that comes from the charged lepton sector.
Finally, we note that the residual flavour symmetry forbids the
Majorana mass terms at any order and provides, by construction, a
natural realization of a type II Dirac see-saw mechanism for small
neutrino masses.

\begin{acknowledgments}

  This research is supported by the Spanish grants FPA2014-58183-P,
  SEV-2014-0398 (MINECO) and PROMETEOII/2014/084 (Generalitat
  Valenciana).  E.P. is supported by DGAPA-PAPIIT RA101516 and
  IN100217.  J.M.L.  would like to thanks AHEP group at IFIC for its
  kind hospitality while part of this work was carried out and
  DGAPA-PAPIIT RA101516 and CONACYT (M\'exico) for finantial support.

\end{acknowledgments}
\appendix

\section{The $A_4$ product representation}
\label{a4rep}

The non abelian discrete group $A_4$, or the group of the even
permutation of four elements, has four irreducible
representations~\cite{Ma:2001dn,Babu:2002dz,King:2014nza}:
three singlets $\bf{1_1}$, $\bf{1_2}$, and $\bf{1_3}$ and one triplet
$\bf{3}$ and two generators: $S$ and $T$ following the relations
$S^2 = T^3 = (ST)^3 = \mathcal{I}$.  The one-dimensional unitary
representations are
\begin{equation}
\begin{array}{lll}
\bf{1_1}:  &S=1, &T=1,\\
\bf{1_2}:  &S=1, &T=\omega,\\
\bf{1_3}:  &S=1, &T=\omega^2,
\end{array}
\end{equation}
where $\omega^3=1$.
In the basis where $S$ is real diagonal,
\begin{equation}
\label{eq:ST}
S=\left(
\begin{array}{ccc}
1&0&0\\
0&-1&0\\
0&0&-1\\
\end{array}
\right)\,\text{and}\quad
T=\left(
\begin{array}{ccc}
0&1&0\\
0&0&1\\
1&0&0\\
\end{array}
\right)\,. 
\end{equation}

The product rule for the singlets are
\begin{equation}
\begin{array}{l}
\bf{1_1}\otimes\bf{1_1} = \bf{1_2} \otimes \bf{1_3} = 1,\\
\bf{1_2}\otimes\bf{1_2} = \bf{1_3},\\
\bf{1_3}\otimes\bf{1_3} = \bf{1_2},
\end{array}
\end{equation} 
and triplet multiplication rules are
\begin{equation}\label{pr}
\begin{array}{lll}
(ab)_{1_1}&=&a_1b_1+a_2b_2+a_3b_3\,,\\
(ab)_{1_2}&=&a_1b_1+\omega a_2b_2+\omega^2a_3b_3\,,\\
(ab)_{1_3}&=&a_1b_1+\omega^2 a_2b_2+\omega a_3b_3\,,\\
(ab)_{3_1}&=&(a_2b_3,a_3b_1,a_1b_2)\,,\\
(ab)_{3_2}&=&(a_3b_2,a_1b_3,a_2b_1)\,,
\end{array}
\end{equation}
where $a=(a_1,a_2,a_3)$ and $b=(b_1,b_2,b_3)$.

\bibliographystyle{bib_style_T1}
\bibliography{S0370269317306214.bib,corfu-2017,newrefs,merged_Valle}  
\end{document}